\documentclass[aps,prl,showpacs,twocolumn]{revtex4}
\usepackage{amsmath}
\usepackage{graphicx}
\usepackage{dsfont}

\begin{document}

\title{SU(3) quantum critical model emerging from a spin-1 topological phase}
\author{Wen-Jia Rao$^{1}$, Guo-Yi Zhu$^{1}$, and Guang-Ming Zhang$^{1,2}$}
\email{gmzhang@tsinghua.edu.cn}
\affiliation{$^{1}$State Key Laboratory of Low-Dimensional Quantum Physics and Department
of Physics, Tsinghua University, Beijing 100084, China. \\
$^{2}$Collaborative Innovation Center of Quantum Matter, Beijing 100084,
China.}
\date{\today}

\begin{abstract}
Different from the spin-1 Haldane gapped phase, we propose a novel SO(3)
spin-1 matrix product state (MPS), whose parent Hamiltonian includes
three-site spin interactions. From the entanglement spectrum of a single
block with $l$ sites, an enlarged SU(3) symmetry is identified in the edge
states, which are conjugate to each other for the $l=even$ block but
identical for the $l=odd$ block. By blocking this novel state, the blocked
MPS explicitly displays the SU(3) symmetry with two distinct structures.
Under a symmetric bulk bipartition with a sufficient large block length $%
l=even$, the entanglement Hamiltonian (EH) of the reduced system
characterizes a spontaneous dimerized phase with two-fold degeneracy.
However, for the block length $l=odd$, the corresponding EH represents an
SU(3) quantum critical point with delocalized edge quasiparticles, and the
critical field theory is described by the SU(3) level-1 Wess-Zumino-Witten
conformal field theory.
\end{abstract}

\maketitle

Topological phases of matter have become one of the most important subjects
in physics, because their low-energy excitations have potential use for
fault-tolerant quantum computation. Symmetry protected topological (SPT)
phases belong to a new type of topological phases with robust gapless edge
excitations\cite{Chen-Gu-Wen-2011,Schuch,chen-gu-liu-wen}. Without breaking
the protecting symmetry or closing the energy gap, these SPT phases can not
be continuously connected to the trivial phase. A topological quantum
critical point (TQCP) exists to separate an SPT phase from its adjacent
trivial phase, and the corresponding critical theory is beyond the
Landau-Ginzburg-Wilson paradigm\cite{Chen-Wang-Lu-Lee,Tsui-Jiang-Lu-Lee}.
The simplest example of SPT phases is the Haldane gapped phase of the
antiferromagnetic Heisenberg spin-1 chain\cite{Haldane-1983}. Recently it
has been shown that the symmetric bulk bipartition of the SO(3) symmetric
Affleck-Kennedy-Lieb-Tasaki (AKLT) wave function for the Haldane phase\cite%
{AKLT} is an effective way to create an array of fractionalized spin-1/2
edge spins in the bulk subsystem, and the corresponding bulk entanglement
spectrum (ES) represents a TQCP described by the SU(2) level-1
Wess-Zumino-Witten (WZW) theory with spinon excitations\cite%
{rao-wan-zhang,rao-zhang-yang}. The TQCP is argued to characterize the
quantum critical state between the Haldane gapped phase and its adjacent
trivial phase\cite{rao-wan-zhang,rao-zhang-yang,hsieh-fu-qi}.

It is well-known that the quantum spin-1 chain may exhibit an SU(3)
symmetry. The SU(3) symmetric matrix product state (MPS) has been constructed%
\cite{Greiter07,Katsura,Rachel,Quella12,Furusaki,Quella15}. The on-site
physical space is spanned by the SU(3) adjoint representation $8$ consisting
of a fundamental representation $3$ (quarks) and a conjugate representation $%
\overline{3}$ (antiquarks), while the adjacent lattice sites are connected
by the SU(3) quark-antiquark singlet bonds. Different from the Haldane
gapped phase, the zero-energy states localized on the edges are conjugate to
each other. The natural question arises whether a distinct spin-1 MPS can be
found so that the symmetric bulk bipartition leads to the SU(3) quantum
criticality for the reduced bulk system. According to the complete
classification\cite{Chen-Gu-Wen-2011,Schuch,chen-gu-liu-wen}, such a
nontrivial topological phase has to be constructed under a projective group
of the SU(3) symmetry.

In this paper, we construct a novel SO(3) symmetric spin-1 MPS with virtual
spin-1 triplets, and the determined parent Hamiltonian includes three-site
spin interactions. From the ES of a single block with $l$ sites, the edge
states exhibit an enlarged SU(3) symmetry: the $l=even$ block with conjugate
edge states while the $l=odd$ block with the same edges. By blocking this
novel SO(3) state, the blocked MPS explicitly displays the SU(3) symmetry
with two distinct structures. Under a symmetric bulk bipartition with a
sufficiently large block length $l=even$, the entanglement Hamiltonian (EH)
of the reduced system describes a spontaneous dimerized phase with two-fold
degeneracy\cite{Barber89,Klumper89}. Such a state may be argued to describe
the first-order phase transition between two SU(3) symmetric
valence-bond-solid states with conjugate edge states. However, for the block
length $l=odd$, the corresponding EH represents an SU(3) quantum critical
state with delocalized edge quasiparticles, characterizing the TQCP between
the spin-1 MPS with the same edge states and the trivial phase. The critical
field theory is described by the 1+1 (space-time) SU(3) level-1 WZW
conformal field theory\cite{Fath,Itoi,Lauchli}.

\textit{SO(3) spin-1 MPS.- }The novel SO(3) symmetric spin-$1$ MPS can be
constructed as follows. We assume that each physical site consists of two
virtual spin-$1$ triplets which are projected into a total spin-$1$ triplet,
while the neighboring sites are linked by the spin-$1$ singlet bonds. Such a
state is different from the AKLT types of MPS, where the on-site virtual
spins are projected into the maximal total spin multiplets. The ground state
wave function for a closed chain is expressed as%
\begin{equation}
|\text{VBS}\rangle =\sum_{\left\{ s_{i}\right\} }\text{Tr}\left( A^{\left[
s_{i}\right] }A^{\left[ s_{2}\right] }...A^{\left[ s_{N}\right] }\right)
|s_{1},s_{2},...s_{N}\rangle ,
\end{equation}%
where the local matrices are given by
\begin{equation*}
A^{\left[ -1\right] }=S^{-}/\sqrt{2},A^{\left[ 0\right] }=S^{z}/\sqrt{2},A^{%
\left[ 1\right] }=-S^{+}/\sqrt{2},
\end{equation*}
with $S^{\pm }=\left( S^{x}\pm iS^{y}\right) /\sqrt{2}$, and $S^{\alpha }$ ($%
\alpha =x,y,z$) denote the spin-1 SO(3) matrices. From the transfer matrix $%
T=\sum_{s}A^{\left[ s\right] }\otimes \bar{A}^{\left[ s\right] }$, the
spin-spin correlation functions are proved to be exponentially decaying with
the correlation length $\xi =1/\ln 2$, which is larger than that of the spin-%
$1$ AKLT state ($1/\ln 3$). For a sufficient long chain with open boundary
condition, there is an energy gap in the low-energy excitations and the
degenerate zero-energy states given by two spin-1 edge triplets. Actually
the existence of such an SO(3) spin-1 MPS had been noticed before\cite%
{Tu09,Jian}, however, its properties have not been explored yet.

The first important thing is to find out the parent Hamiltonian. According
to the tensor product of two spin-1's: $1\otimes 1=0\oplus 1\oplus 2$, there
are no null states in the Hilbert space of two sites. So we have to consider
the spin interactions among the spin-1 operators ($s_{1}$,$s_{2}$,$s_{3}$)
on three successive sites. Under the site-centered inversion symmetry, we
first couple $s_{1}$ and $s_{3}$ and then with $s_{2}$. The resulting spin
multiplets are labeled by $|S_{13};S,M\rangle $, where $S_{13}$ and $S$ are
good quantum numbers and $M=-S,..,S$. By means of the Clebsch-Gordan
coefficients $C_{S_{13}}^{\left[ S,M\right] }$, the local three-spin state
can be expanded into {\footnotesize
\begin{eqnarray}
&&\sum_{\left\{ s_{i}\right\} }A^{\left[ s_{1}\right] }A^{\left[ s_{2}\right]
}A^{\left[ s_{3}\right] }|s_{1}s_{2}s_{3}\rangle  \notag \\
&=&-\frac{1}{2\sqrt{3}}\left| 1;0,0\right\rangle +\sum_{M}C_{1}^{\left[ 2,M%
\right] }\left| 1;2,M\right\rangle  \notag \\
&&+\sum_{M}C_{0}^{\left[ 1,M\right] }\left( \left| 0;1,M\right\rangle +\sqrt{%
5}\left| 2;1,M\right\rangle \right) ,
\end{eqnarray}%
} indicating that the relevant spin multiplets correspond to total spin
channels $S=0,1,2$, consistent with the result of two spin-$1$ edge
triplets. The parent Hamiltonian is given by the null states which are
absent in the above blocking states, $H=\sum_{i}h_{i}$, where{\footnotesize
\begin{eqnarray}
&&h_{i}=\lambda _{1}\sum_{M}\left( \sqrt{5}|0;1,M\rangle -|2;1,M\rangle
\right) \left( \sqrt{5}\langle 0;1,M|-\langle 2;1,M|\right)  \notag \\
&&\text{ \ \ }+\lambda _{2}\sum_{M}|2;2,M\rangle \langle 2;2,M|+\lambda
_{3}\sum_{M}|2;3,M\rangle \langle 2;3,M|  \notag \\
&&\text{ \ \ }+\lambda _{4}\sum_{M}|1;1,M\rangle \langle 1;1,M|,
\end{eqnarray}%
} with all positive coefficients $\lambda _{j}$. When the above projections
are expressed in terms of the physical spin-1 operators, the local parent
Hamiltonian becomes
\begin{eqnarray}
&&h_{i}=a_{0}+a_{1}\left( \mathbf{s_{i-1}}\cdot \mathbf{s_{i}}+\mathbf{s_{i}}%
\cdot \mathbf{s_{i+1}}\right) +a_{2}\mathbf{s_{i-1}}\cdot \mathbf{s_{i+1}}
\notag \\
&&\text{ \ \ }+a_{3}\left[ (\mathbf{s_{i-1}}\cdot \mathbf{s_{i}})^{2}+(%
\mathbf{s_{i}}\cdot \mathbf{s_{i+1}})^{2}\right] +a_{4}(\mathbf{s_{i-1}}%
\cdot \mathbf{s_{i+1}})^{2}  \notag \\
&&\text{ \ \ }+a_{5}\left[ (\mathbf{s_{i-1}}\cdot \mathbf{s_{i}})(\mathbf{%
s_{i}}\cdot \mathbf{s_{i+1}})+(\mathbf{s_{i+1}}\cdot \mathbf{s_{i}})(\mathbf{%
s_{i}}\cdot \mathbf{s_{i-1}})\right]  \notag \\
&&\text{ \ \ }+a_{6}\left( \mathbf{s_{i-1}}\cdot \mathbf{s_{i}}+\mathbf{s_{i}%
}\cdot \mathbf{s_{i+1}}\right) (\mathbf{s_{i-1}}\cdot \mathbf{s_{i+1}})
\notag \\
&&\text{ \ \ }+a_{7}\left[ (\mathbf{s_{i-1}}\cdot \mathbf{s_{i}})(\mathbf{%
s_{i-1}}\cdot \mathbf{s_{i+1}})(\mathbf{s_{i}}\cdot \mathbf{s_{i+1}})\right.
\notag \\
&&\text{ \ \ \ \ }+\left. (\mathbf{s_{i}}\cdot \mathbf{s_{i+1}})(\mathbf{%
s_{i-1}}\cdot \mathbf{s_{i+1}})(\mathbf{s_{i}}\cdot \mathbf{s_{i-1}})\right]
,
\end{eqnarray}%
with the coupling parameters
\begin{eqnarray}
a_{0} &=&\left( 15\lambda _{1}+5\lambda _{2}+\lambda _{3}+9\lambda
_{4}\right) /15,  \notag \\
a_{1} &=&(\lambda _{3}-\lambda _{2})/6,  \notag \\
a_{2} &=&(10\lambda _{2}+2\lambda _{3}-10\lambda _{1}-7\lambda _{4})/20,
\notag \\
a_{3} &=&(4\lambda _{3}-30\lambda _{1}-10\lambda _{2}-39\lambda _{4})/120,
\notag \\
a_{4} &=&(10\lambda _{2}+2\lambda _{3}-30\lambda _{1}+3\lambda _{4})/60,
\notag \\
a_{5} &=&(2\lambda _{3}+3\lambda _{4}-5\lambda _{2})/30,  \notag \\
a_{6} &=&(\lambda _{3}-\lambda _{4})/10,  \notag \\
a_{7} &=&(30\lambda _{1}+4\lambda _{3}+21\lambda _{4}-10\lambda _{2})/120.
\end{eqnarray}
The three-site spin interactions are involved! Actually, the most important
feature is the spin-1 edge triplets, but we do not know whether these edge
states are symmetry protected or not.

\textit{Hidden symmetry in the edge states.- }In order to reveal the hidden
symmetry in the edge spin-1 triplets, we study the entanglement properties
of a single block. To this end, it is convenient to introduce
three-component fermions to represent the \textit{physical} $s=1$ triplet
\begin{equation}
|1\rangle =c_{1}^{\dagger }c_{0}^{\dagger }|vac\rangle ,|0\rangle
=c_{1}^{\dagger }c_{-1}^{\dagger }|vac\rangle ,|-1\rangle =c_{0}^{\dagger
}c_{-1}^{\dagger }|vac\rangle ,
\end{equation}%
and the SO(3) spin-1 operators are expressed as $s_{i}^{\alpha }=\sum_{\mu
,\nu }c_{i\mu }^{\dagger }S_{\mu \nu }^{\alpha }c_{i\nu }$. In terms of
these fermions, the spin-$1$ MPS can be written as the product of SO(3) bond
singlets.
\begin{equation}
|\text{VBS}\rangle =\prod_{i}\left( c_{i,1}^{\dagger }c_{i+1,-1}^{\dagger
}-c_{i,0}^{\dagger }c_{i+1,0}^{\dagger }+c_{i,-1}^{\dagger
}c_{i+1,1}^{\dagger }\right) |vac\rangle .
\end{equation}

For a quantum spin-1, spin quadruple operators can be defined by
\begin{eqnarray}
Q^{1} &=&\left( S^{x}\right) ^{2}-\left( S^{y}\right) ^{2},Q^{2}=\left[
3\left( S^{z}\right) ^{2}-2\right] /\sqrt{3},  \notag \\
Q^{3} &=&S^{y}S^{z}+S^{z}S^{y},Q^{4}=S^{z}S^{x}+S^{x}S^{z},  \notag \\
Q^{5} &=&S^{x}S^{y}+S^{y}S^{x},
\end{eqnarray}
and then total nine spin multiplets (singlet, triplet, and quintet) are
formed by two virtual spin-1 states. The corresponding wave functions in
terms of a Nambu spinor $\varphi _{j}=\left( c_{j,1},c_{j,0},c_{j,-1}\right)
^{t}$ are expressed as
\begin{eqnarray}
|\psi ^{-}\rangle _{j,j+1} &=&\frac{1}{\sqrt{3}}\varphi _{j}^{\dagger
}R\left( \varphi _{j+1}^{\dagger }\right) ^{t}|vac\rangle ,  \notag \\
|\psi ^{\alpha }\rangle _{j,j+1} &=&\frac{1}{\sqrt{2}}\varphi _{j}^{\dagger
}M^{\alpha }\left( \varphi _{j+1}^{\dagger }\right) ^{t}|vac\rangle ,
\end{eqnarray}%
where $\alpha =1,.,8$, $R=1-2\left( S^{y}\right) ^{2}$, $M^{1}=-Q^{5}$, $%
M^{2}=Q^{1}$, $M^{3}=Q^{4}$, $M^{4}=Q^{3}$, $M^{5}=Q^{2}$, $M^{6}=-iS^{y}$, $%
M^{7}=-iS^{x}$, and $M^{8}=iS^{z}$\textbf{.} The MPS wave function is
further transformed into
\begin{eqnarray}
|\text{VBS}\rangle &=&\frac{1}{\sqrt{3}}\frac{1}{2^{N/2}}\sum_{\left\{
\alpha _{i}=6,7,8\right\} }|\psi ^{\alpha _{1}}\rangle |\psi ^{\alpha
_{2}}\rangle ..|\psi ^{\alpha _{N}}\rangle  \notag \\
&&\times \varphi _{0}^{\dagger }M^{\alpha _{1}}M^{\alpha _{2}}..M^{\alpha
_{N}}R\left( \varphi _{N+1}^{\dagger }\right) ^{t}|vac\rangle ,
\end{eqnarray}%
where we have introduced the boundary Nambu spinors $\varphi _{0}^{\dagger }$
and $\varphi _{N+1}^{\dagger }$ to fix the edge spins.

\begin{figure}[t]
\includegraphics[width=8cm]{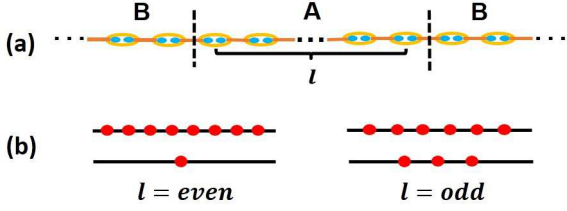} % Here is how to import EPS art
\caption{(a) The novel SO(3) symmetric spin-1 MPS. Each blue dot represents
a virtual spin-1, yellow circles stand for local spin-1 triplets, and solid
lines denote the singlet bonds. A block with the length $l$ is chosen as the
subsystem A. (b) The entanglement spectrum of the block with even $l$ (left)
and odd $l$ (right).}
\label{fig:ESOneBlock}
\end{figure}

When a block with length $l\geq 2$ is picked out as the subsystem A shown in
Fig.\ref{fig:ESOneBlock}(a), we can trace out the degrees of freedom of the
other subsystem and the reduced density matrix is derived%
\begin{eqnarray}
\rho _{A} &=&\frac{1}{2^{l}}\sum_{\left\{ \alpha _{i},\alpha _{i}^{\prime
}\right\} }\frac{1}{9}\text{Tr}\left( \varphi _{0}^{\dagger }V^{\prime
\dagger }V\varphi _{0}\right) _{0,N+1}  \notag \\
&&\times |\psi ^{\alpha _{1}}\rangle \langle \psi ^{\alpha _{1}^{\prime
}}|...|\psi ^{\alpha _{l}}\rangle \langle \psi ^{\alpha _{l}^{\prime }}|,
\end{eqnarray}%
where $\alpha _{i},\alpha _{i}^{\prime }=6,7,8$, $V=M^{a_{1}}...M^{\alpha
_{l}}$ and $V^{\prime }=M^{\alpha _{1}^{\prime }}...M^{\alpha _{l}^{\prime
}} $. Generally there are nine degenerate states with different edge states
for the subsystem A. When the reduced density matrix $\rho _{A}$ is applied
to these states, the corresponding eigenvalues are derived {\footnotesize
\begin{eqnarray}
\lambda _{s} &=&\frac{1}{9}\left[ 1+3\left( \frac{1}{2}\right) ^{l}+5\left( -%
\frac{1}{2}\right) ^{l}\right] ,  \notag \\
\lambda _{t} &=&\frac{1}{9}\left[ 1+\frac{3}{2}\left( \frac{1}{2}\right)
^{l}-\frac{5}{2}\left( -\frac{1}{2}\right) ^{l}\right] ,  \notag \\
\lambda _{q} &=&\frac{1}{9}\left[ 1-\frac{3}{2}\left( \frac{1}{2}\right)
^{l}+\frac{1}{2}\left( -\frac{1}{2}\right) ^{l}\right] ,
\end{eqnarray}%
}for the SO(3) singlet, triplet, and quintet, respectively. However, it is
more interesting to notice that the eigenvalues of the triplet and quintet
become identical for $l=even$, while the eigenvalues of the singlet and
quintet are the same for $l=odd$. The single block ES defined by $\xi
_{i}=-\ln \lambda _{i}$ is displayed in Fig.\ref{fig:ESOneBlock}(b).

Corresponding to the ES for $l=even/odd$, the entanglement Hamiltonian (EH),
$H_{E}=-\ln \rho _{A}$, can be characterized by the two edge spins of the
block: $H_{E}^{even}=-\frac{3}{2^{l}}\left( \mathbf{t}_{L}\cdot \mathbf{t}%
_{R}\right) ^{2}$ and $H_{E}^{odd}=\frac{3}{2^{l}}\left[ \left( \mathbf{t}%
_{L}\cdot \mathbf{t}_{R}\right) +\left( \mathbf{t}_{L}\cdot \mathbf{t}%
_{R}\right) ^{2}\right] $. Although the bulk MPS has only the SO(3)
symmetry, the edge states display an enlarged $SU\left( 3\right) $ symmetry.
$H_{E}^{even}$ can be rewritten under the SU(3) fundamental representation $%
3 $ (quarks) and its conjugate representation $\overline{3}$ (antiquarks)
for each edge spin ($3\otimes \overline{3}=1\oplus 8$), describing an SU(3)
singlet and an octet in the ES for $l=even$. On the other hand, $H_{E}^{odd}$
can be expressed under either the SU(3) fundamental or conjugate
representation for both edge spins ($3\otimes 3=\bar{3}\oplus 6$ or $\bar{3}%
\otimes \bar{3}=3\oplus \bar{6}$), yielding an SU(3) triplet and a sextet in
the ES for $l=odd$. So the two edges are \textit{conjugate} with each other
for $l=even$, while they are the \textit{same} for $l=odd$. It should be
pointed out that the ES of a single block is just identical to the edge
excitation spectrum for an open spin chain. The spin-1 edge triplets are
thus protected by the enlarged SU(3) symmetry.

\begin{figure}[t]
\includegraphics[width=8cm]{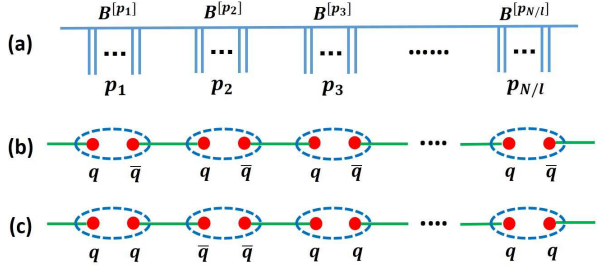} % Here is how to import EPS art
\caption{(a) The blocked MPS wave function with the block length $l$. (b)
For even $l$, each block contains a pair of SU(3) quark and antiquark. (c)
For odd $l$, a pair of quarks is included in the odd number blocks while a
pair of antiquarks in the even number blocks.}
\label{fig:BlockMPS}
\end{figure}

\textit{SU(3) symmetry in the bulk.- }To make the hidden structure of this
spin-$1$ SPT state explicitly, we perform the exact state renormalization.
By applying the singular value decomposition to a block with the length $l$,
\begin{equation}
\left( A^{\left[ s_{1}\right] }A^{\left[ s_{2}\right] }..A^{\left[ s_{l}%
\right] }\right) _{\alpha ,\beta }=X_{\left\{ s_{i}\right\} ,p}\Lambda
_{p,p}Y_{p,\left( \alpha ,\beta \right) },
\end{equation}%
we find nine relevant states $|p\rangle $, which can be viewed as the
resulting states of two spin-$1$ edge states. When these relevant states are
chosen as the basis of each block, the original MPS can be written into%
\begin{equation*}
|\text{VBS}\rangle =\sum_{\left\{ p_{i}\right\} }\text{Tr}\left( B^{\left[
p_{1}\right] }B^{\left[ p_{2}\right] }...B^{\left[ p_{N/l}\right] }\right)
|p_{1},p_{2},...p_{N/l}\rangle ,
\end{equation*}%
where the block matrices are expressed in terms of the generators of the
SU(3) group {\footnotesize
\begin{eqnarray*}
B^{[0]} &=&\sqrt{\lambda _{s}}\mathds{1},\text{ }B^{[1]}=\sqrt{3\lambda
_{t}/2}S^{-},\text{ }B^{[2]}=\sqrt{3\lambda _{t}/2}S^{z}, \\
B^{[3]} &=&-\sqrt{3\lambda _{t}/2}S^{+},\text{ }B^{[4]}=\sqrt{3\lambda _{q}/4%
}\left( Q^{1}-iQ^{5}\right) , \\
B^{[5]} &=&\sqrt{3\lambda _{q}/4}\left( Q^{4}-iQ^{3}\right) ,\text{ }B^{[6]}=%
\sqrt{3\lambda _{q}/2}Q^{2}, \\
B^{[7]} &=&-\sqrt{3\lambda _{q}/4}\left( Q^{4}+iQ^{3}\right) ,\text{ }%
B^{[8]}=\sqrt{3\lambda _{q}/4}\left( Q^{1}+iQ^{5}\right) .
\end{eqnarray*}%
} The blocked MPS wave function is displayed in Fig.\ref{fig:BlockMPS}(a).

As pointed out in the analysis of edge states, for the even block length,
the nine local relevant states $|p\rangle $ are represented by a pair of the
SU(3) quark and antiquark, and the resulting blocked wave function is shown
in Fig.\ref{fig:BlockMPS}(b). Such a state belongs to the same class of the $%
SU\left( 3\right) $ symmetric MPS\cite%
{Greiter07,Katsura,Rachel,Quella12,Furusaki,Quella15}. However, for the odd
block length, the nine relevant states $|p\rangle $ are represented by two
SU(3) quarks $3$ in the odd number blocks and two antiquarks $\overline{3}$
in the even number of blocks. Such an SU(3) symmetric blocked MPS has a
doubled lattice unit cell, shown in Fig.\ref{fig:BlockMPS}(c).

Combining with the analysis of edge states, we can further reveal the hidden
structure in the original SO(3) MPS wave function. Under the SU(3)
representation, the original lattice is divided into two sublattices. On the
odd lattice sites there are two fundamental SU(3) quark states and projected
into a conjugate SU(3) triplet $\overline{3}$, while two conjugate SU(3)
antiquark states on the even lattice sites which are projected into a
fundamental SU(3) triplet $3$. The adjacent sites are linked by the SU(3)
quark-antiquark singlet bonds. For an open chain with the total length $L$,
two distinct topological states can be classified. For $L=even$, the
representations of two edges are conjugate to each other, while they are
identical for $L=odd$. The former represents the SU(3) valence-bond-solid
state, and the latter stands for a novel SU(3) spin-1 SPT phase. To our
knowledge, the novel spin-1 MPS has not been found before.

\begin{figure}[t]
\includegraphics[width=8cm]{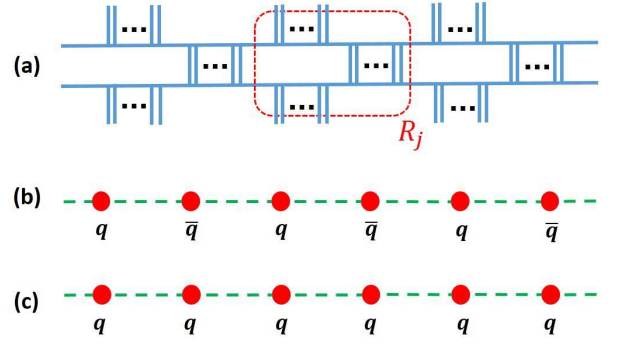} % Here is how to import EPS art
\caption{(a) The reduced density matrix under symmetric bulk bipartition
with the repeating structure $R_{j}$. (b) For the block length $l=even$, the
EH describes an array of alternating SU(3) quarks and antiquarks with the
nearest neighbor interactions. (c) For the block length $l=odd$, the EH
characterizes an array of quarks with the nearest neighbor interactions.}
\label{fig:MPO}
\end{figure}

\textit{Symmetric bulk bipartition for the blocked MPS.- }To extract the
quantum critical properties encoded in this topological phase, we perform
the symmetric bulk bipartition\cite{rao-wan-zhang,rao-zhang-yang}. When the
blocked lattice is divided into two subsystems both containing the same
number of disjoint blocks, we trace out the degrees of freedom in the
sublattice B, i.e., a collection of all even blocks. Then the reduced
density matrix $\rho _{A}$ can be derived in terms of blocked transfer
matrix $T_{b}=\sum_{p}B^{\left[ p\right] }\bar{B}^{\left[ p\right] }$. The
full expression for $\rho _{A}$ shown in Fig.\ref{fig:MPO}(a) can be written
into the matrix product form $\rho _{A}=$Tr$\left( \prod_{j}R_{j}\right) $.

Since the original MPS is gapped with short-range spin correlations, the
couplings between the edge spins within one block and adjacent disjoint
blocks are almost identical for a sufficient large block length $l$. To the
first order of with $\varepsilon =\left( 1/2\right) ^{l}$ and $\delta
=\left( -1/2\right) ^{l}$, $R_{j}$ can be separated into $R_{j}=\frac{1}{6}%
\widetilde{R}_{2j-1}\widetilde{R}_{2j}$, where {\tiny
\begin{equation*}
\widetilde{R}_{j}=\left(
\begin{array}{ccccccccc}
\sqrt{\frac{2}{3}} & \varepsilon S_{j}^{z} & -\varepsilon S_{j}^{x} &
-i\varepsilon S_{j}^{y} & \delta Q_{j}^{1} & \sqrt{\frac{5}{3}}\delta
Q_{j}^{2} & -\delta Q_{j}^{4} & i\delta Q_{j}^{5} & -i\delta Q_{j}^{3} \\
-S_{j}^{z} & 0 & 0 & 0 & 0 & 0 & 0 & 0 & 0 \\
S_{j}^{x} & 0 & 0 & 0 & 0 & 0 & 0 & 0 & 0 \\
-iS_{j}^{y} & 0 & 0 & 0 & 0 & 0 & 0 & 0 & 0 \\
Q_{j}^{1} & 0 & 0 & 0 & 0 & 0 & 0 & 0 & 0 \\
\sqrt{\frac{3}{5}}Q_{j}^{2} & 0 & 0 & 0 & 0 & 0 & 0 & 0 & 0 \\
-Q_{j}^{4} & 0 & 0 & 0 & 0 & 0 & 0 & 0 & 0 \\
-iQ_{j}^{5} & 0 & 0 & 0 & 0 & 0 & 0 & 0 & 0 \\
iQ_{j}^{3} & 0 & 0 & 0 & 0 & 0 & 0 & 0 & 0%
\end{array}%
\right) .
\end{equation*}%
} Then the corresponding EH can be derived accordingly.

\begin{figure}[t]
\includegraphics[width=8cm]{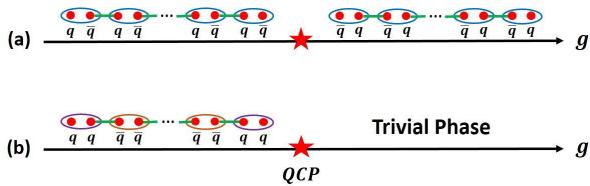}
\caption{The phase diagrams are suggested as the controlling parameter $g$.
(a) The first order phase transition between two SU(3) valence-bond-solid
states. (b) The quantum critical point exists between the SU(3) topological
nontrivial phase with identical edges and the trivial phase.}
\label{fig:QCP}
\end{figure}

For the block length $l=even$, the resulting EH is approximated by
\begin{equation}
H_{E}\simeq -3\left( \frac{1}{2}\right) ^{l}\sum_{j}\left( \mathbf{t}%
_{j}\cdot \mathbf{t}_{j+1}\right) ^{2},
\end{equation}%
where the spin exchange interactions transform under the SU(3) fundamental
representation $3$ and the conjugate representation $\overline{3}$ on every
other sites, as displayed in Fig.\ref{fig:MPO}(b). Since this model is
exactly solvable, the ground state is given by a gapped dimerized phase with
two-fold degeneracy\cite{Barber89,Klumper89}. In the thermodynamic limit,
the normalized energy gap is $0.1732$, the spin correlation length is $\xi
_{\lambda }\simeq 21.073$, and the translational invariance has been
spontaneously broken. Actually, such a state can be regarded as the
quark-antiquark singlet bonds of the edge states for each block, and the
spontaneously breaking of the bond-centered inversion symmetry gives rise to
the two-fold degeneracy. So the EH can be regarded to describe the
first-order phase transition between two SU(3) symmetric valence-bond-solid
states shown in Fig.\ref{fig:QCP}(a).

For the block length $l=odd$, the corresponding EH can be derived as
\begin{equation}
H_{E}\simeq 3\left( \frac{1}{2}\right) ^{l}\sum_{j}\left[ \left( \mathbf{t}%
_{j}\cdot \mathbf{t}_{j+1}\right) +\left( \mathbf{t}_{j}\cdot \mathbf{t}%
_{j+1}\right) ^{2}\right] ,
\end{equation}%
where the spin exchange interactions transform under either the SU(3)
fundamental representation $3$ or the conjugate representation $\overline{3}$
on every sublattice site, shown in Fig.\ref{fig:MPO}(c). Actually, $H_{E}$
is also an exactly solvable model, corresponding to an SU(3) symmetric
quantum critical state\cite{Fath,Itoi,Lauchli}. The corresponding critical
field theory is described by the 1+1 (space-time) dimensional $SU\left(
3\right) _{1}$\ WZW conformal field theory with the central charge\emph{\ }$%
c=2$. The primary fields are given by the quantum spin numbers $j=0,1,2$
with the scaling dimensions $\Delta _{0}=0,\Delta _{1}=1/3,$and $\Delta
_{2}=1$, respectively. The low-energy elementary excitations have the same
nature of the edge states of the blocked MPS. Such a critical state can be
argued to characterize the TQCP between the spin-1 SPT with identical edges
and the trivial gapped phase, as shown in Fig.\ref{fig:QCP}(b).

\textit{Conclusion and Outlook.-} A novel SO(3) symmetric spin-1 MPS has
been constructed and its parent Hamiltonian is determined to include
three-site interactions. The peculiar property of this wave function is
that, for an open chain with the total length $L=even$, the representations
of two edges are conjugate to each other, while they are identical for the
system length $L=odd$. Under the SU(3) representation, the orignal MPS can
be transformed as follows: two fundamental SU(3) quark states on the odd
lattice sites are projected into a conjugate SU(3) triplet $\overline{3}$,
while two conjugate SU(3) antiquark states on the even lattice sites are
projected into a fundamental SU(3) triplet $3$. The adjacent sites are
linked by the SU(3) quark-antiquark singlet bonds.

With the same strategies, we can construct a generalized novel MPS with
hidden $SU(N)$ symmetry ($N\geq 3$) with two virtual $S=\left( N-1\right) /2$
spins which are projected into the sum of antisymmetric channels. For odd $N$%
, the antisymmetric channels consist of the total odd spin multiplets, while
for even $N$ they are composed of the total even spin multiples. Under the
symmetric bulk bipartition with the block length $l=even$, we have obtained
the EH for the reduced system, which describes a gapped dimerized phase with
two-fold degenerate ground state. For the block length $l=odd$, the
corresponding EH characterizes the $SU(N)$ quantum critical state with
delocalized edge quasiparticles. We will report these results in our future
work.

\textit{Acknowledgements.- } {G. M. Zhang would like to thank D. H. Lee for
the stimulating discussions and acknowledge the support of NSF-China through
Grant No.20121302227.}

\end{document}